# Magnetotransport in wide parabolic PbTe quantum wells

J. Oswald, G. Heigl, M. Pippan, G. Span, T. Stellberger

Institut für Physik, Montanuniversität Leoben, Franz Josef Str. 18, A-8700 Leoben, Austria

D.K. Maude, J.C. Portal

High Magnetic Field Laboratory, CNRS, 25 Avenue des Martyrs, BP 166 Grenoble, France

**Abstract**

The 3D- and 2D- behaviour of wide parabolic PbTe single quantum wells, which consist of PbTe p-n-p-structures, are studied theoretically and experimentally. A simple model combines the 2D- subband levels and the 3D-Landau levels in order to calculate the density of states in a magnetic field perpendicular to the 2D plane. It is shown that at a channel width of about 300nm on can expect to observe 3D- and 2D-behaviour at the same time. Magnetotransport experiments in selectively contacted Hall bar samples are performed at temperatures down to T = 50 mK and at magnetic fields up to B = 17 T.

**Introduction**

Wide parabolic quantum wells (WPQW) are of particular interest because of the intermediate dimensionality of two- (2D) and three- dimensional (3D) physics. During the past few years much effort in fabricating, investigating and modelling [1,2,3] of $Al_xGa_{1-x}As$ - WPQW's have been made. The realization of wide quantum wells with a flat potential in the electron channel requires a parabolic bare potential which is difficult to obtain by MBE growth in quantum well systems. WPQW's can be made much easier by using the nipi-concept [4].

PbTe is a narrow gap semiconductor with both the conduction and the valence band extrema located at the L points of the Brillouin zone. The effective mass ellipsoids are oriented along the four <111> directions and have a mass anisotropy of about a factor of 10. With respect to the growth direction there are two different effective masses for both the subband quantization and the Landau level splitting. Therefore there are two sets of subbands and two sets of Landau levels for a magnetic field perpendicular to the 2D-plane.

The PbTe layer systems are grown on <111> $BaF_2$ substrates by HWE (Hot Wall Epitaxy). The p-layers of the p-n-p structures are designed to be non-depleted in order to screen the embedded n-channel from any influence from surface and interface states [4]. Selective contacts to the embedded n-channel are realised [5] in order to perform magneto transport experiments. The mobility in this selectively conducted structures is in the order of $1 \times 10^5 ... 1 \times 10^6$ $cm^2 V^{-1} s^{-1}$ at lowest temperatures. The large dielectric constant ($\varepsilon$ = 500 - 1000) leads to wide quasi parabolic potential wells with a typical subband separation ranging from 1- 3 meV.

**Model**

The subband energies are calculated self consistently by a numerical solution of the Schrödinger equation within the effective mass approximation at zero magnetic field. The LL's are obtained from a 3D-calculation according to [6,7]. The energy level system $E_{ij}(B)$ consists of all Landau levels (LL's) $E_j(B)$ sitting on all subband levels $E_i$ according to a "rigid potential model". Since in wide quantum wells the Landau level separation is much higher than the subband splitting, the LL's can be considered to be split by the subband quantization. The calculation of the density of states (DOS) is done by a superposition of the individual Landau peaks with a broadening factor $b$. The broadening function in this approximation is assumed to be Gaussian. The DOS as a function of energy $E$ and magnetic field $B$ then reads as follows [8]:

$$DOS(E,B) = \frac{q_e \cdot B}{h} \sum_{i,j} \frac{1}{b \cdot \sqrt{\pi}} \cdot \left( e^{-\frac{(E-E_{ij}^l(B))^2}{b^2}} + 3 \cdot e^{-\frac{(E-E_{ij}^o(B))^2}{b^2}} \right)$$

$l$ denotes the longitudinal and $o$ the oblique valleys, $i$ denotes the subband level index and $j$ the Landau level index. The numerical factor 3 in the second term in brackets accounts for the threefold degeneracy of the oblique valleys.

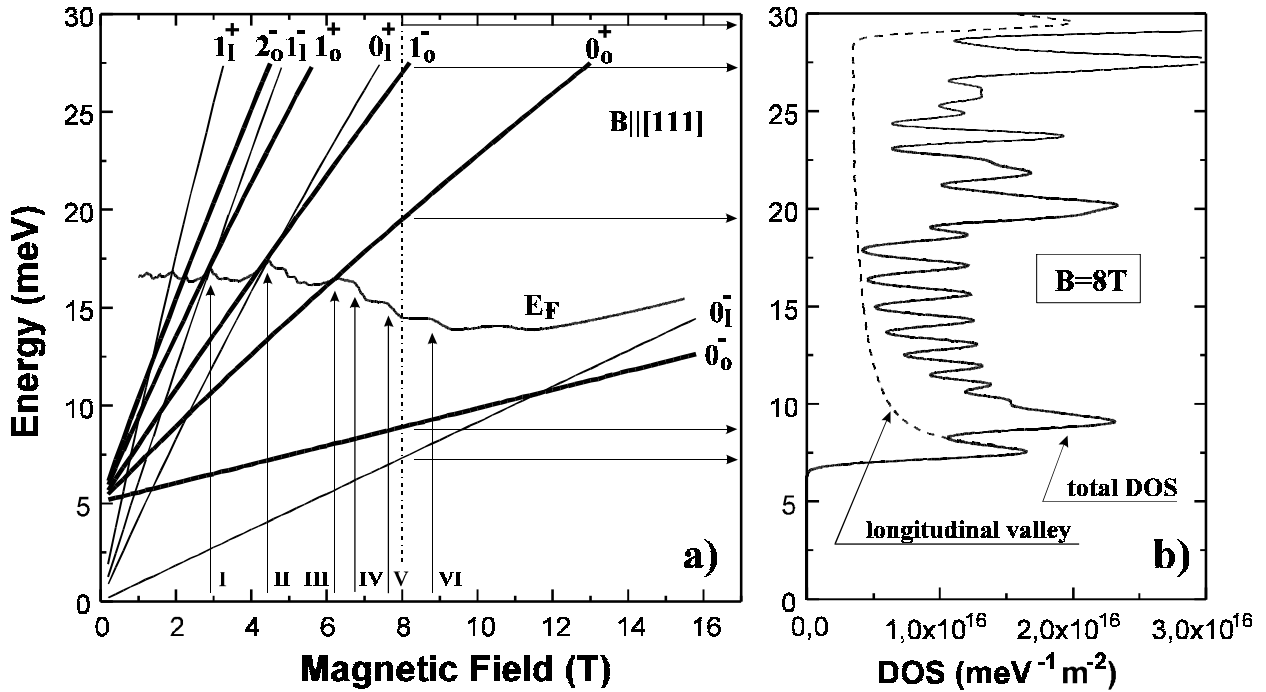

Fig 1a) Landau levels of the subband ground states and Fermi energy versus magnetic field. b) density of states at B = 8 Tesla. Dashed line in b): contribution of the longitudinal valley to the total DOS.

Fig. 1 shows the result of the model calculation of a WPQW with parameters according to the sample used in the experiment. For this sample the carrier sheet density is $n^{2D} = 9.8 \times 10^{12} cm^{-2}$ and the background doping level in the electron channel is $N_D = 3 \times 10^{17} cm^{-3}$ resulting in an effective electron channel width of $d_n = 330$ nm. The shape of the screened potential is a mixture of a square potential (flat region in the middle of the channel) and a parabolic potential. Therefore the subband splitting is very small for the lower subbands and it increases with the subband index. Finally at high subband energies it approaches a constant subband spacing according to the parabolic bare potential. Consequently the associated Landau levels (of same Landau index) of the lower subbands cause an overall enhancement of the density of states near the subband ground state. The energy of the subband ground state is nearly identical to the position of the associated 3D - LL (LL of a three dimensional sample). At higher subband energies the subband levels are separated more and more and therefore a modulation of the DOS by the subband structure is more and more pronounced. For magneto transport experiments the 2D-model predicts an overall enhancement of the DOS at the energy position of the 3D-LL's and therefore also bulk like Shubnikov - de Haas (SdH) oscillations of the magneto resistance can be expected. At higher subband energies, which means the energy range between the 3D-LL's, a modulation of the DOS due to the subband structure of the oblique valleys can be expected. The LL's of the subband ground states are plotted in Fig.1a. The LL-energies at B = 8 T are marked by arrows guiding from Fig.1a to Fig.1b in which the total DOS at B = 8 T is plotted. The contribution of the longitudinal valley to the DOS is a uniform background because of the much smaller subband separation. The contribution of the longitudinal valley is plotted by a dashed line. Due to the lattice mismatch between $BaF_2$ and PbTe a strain induced energy shift between the longitudinal and the oblique valleys of 5 meV is included in the calculation. The energy shift depends strongly on the sample history and is consistent with the value of Ref. [9].

**Experiments**

The sample parameters in the experiment are those used for the model calculation in the previous section. The shape of the sample is a conventional Hall bar of w = 90 μm width. The experiments where performed in an 11 Tesla super conducting magnet at pumped $He^4$ and in a 17 Tesla super conducting magnet with a dilution refrigerator. A conventional lock-in technique using an ac-current of I =10 nA at T = 50 mK and I =100nA at T = 2.2K was used.

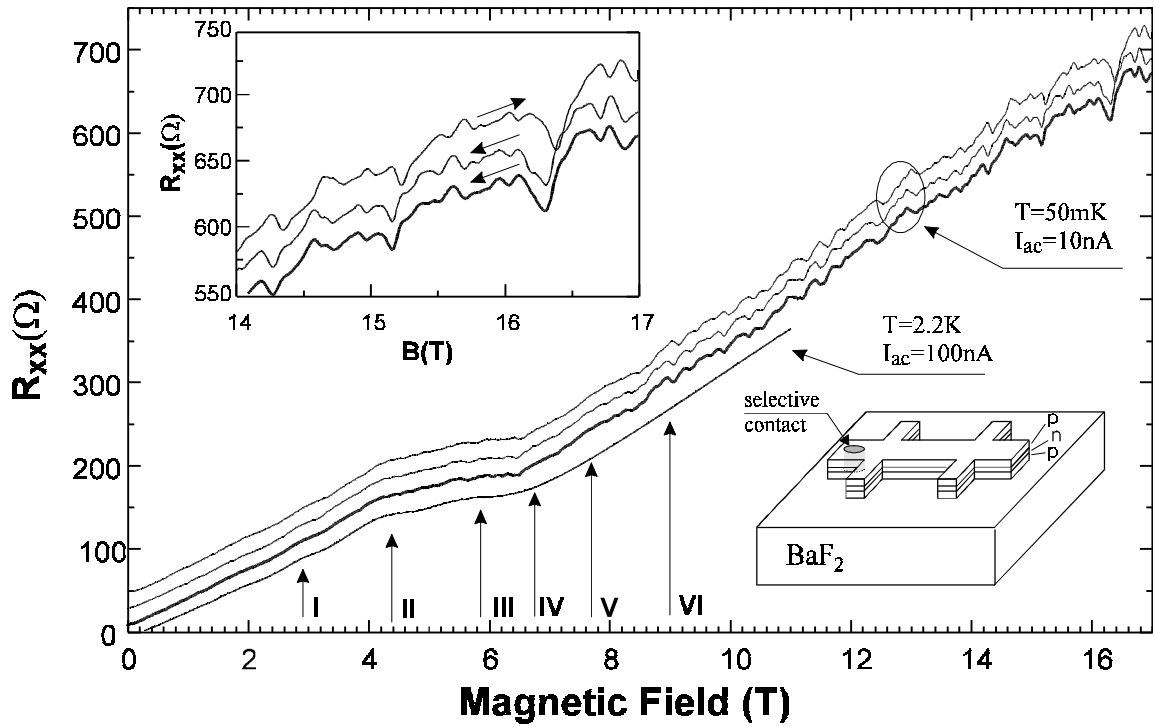

Fig.2: Magneto resistance data. The labelled arrows indicate the position of the main features which have to be compared with Fig.1. Upper insert: magnification of the higher field range. The arrows indicate the sweep direction. Lower insert: Scheme of the sample configuration with one representative selective contact.

Shubnikov - de Haas oscillations according to the 3D parameters of the sample dominate the magnetotransport experiments (Fig.2) and no additional structure in the data at T = 2.2 K is found. At lower temperatures structures in addition to that found at T = 2.2 K appear in the upper field range. This smooth structures are covered by strong fluctuations which are perfectly reproducible. The position of the main features are labelled I, II, III, IV, V, VI and have to be compared with the calculation in Fig.1. Three different sweeps at 50mK are shown in order to demonstrate the reproducibility of the data. For better visibility all plots apart from the thick solid trace are shifted vertically by a small amount.

**Discussion and concluding remarks**

The modulation of the DOS at the Fermi level results from changes of the individual filling factors of the different subband systems involved. The carrier distribution between the different subband systems changes with the magnetic field. Since only the total filling factor is periodic in 1/B a plot of the data over 1/B cannot be used to check the periodicity. At 2.2 K only a structure at I, II and III is visible while structures at IV, V and VI appear at lowest temperatures only. From the comparison with the calculation I, II and III can be attributed to 3D-SdH oscillations and IV, V, VI can be attributed to the 2D subband quantization. The appearance of strong fluctuations in parallel to the appearance of the 2D-features is an indication of the formation of edge states by the oblique valleys while at the same time the longitudinal valley still enables bulk conduction. As an example, an intermediate regime between quantum edge conduction and bulk transport was investigated by Geim et. al. in $n^+$-GaAs epilayers [10] and it was shown that the edge states play an important role in carrying the current to classically inaccessible regions. In contrast to the GaAs system the edge state conduction and the bulk conduction result from electrons in different valleys. Therefore a co-existence of perfect edge state conduction and bulk conduction is possible and new effects resulting from the interplay of these two electron systems can be expected. The observed fluctuations are the subject of further investigations and will be published elsewhere.

**Acknowledgements**

Financial support by Fonds zur Förderung der wissenschaftlichen Forschung (FWF) Vienna (Project No. P10510-NAW) and by Steiermärkischer Wissenschafts- und Forschungslandesfonds.

**References**

[1]    A.J. Rimberg and R.M. Westervelt, Phys. Rev. B **40**, 3970 (1989)
[2]    C.E. Hembre, B.A. Mason, J.T. Kwiatkowski and J.E. Furneaux, Phys. Rev. B **50**, 15197 (1994)
[3]    M. Shayegan, T. Sajoto, M. Santos and C. Silvestre, Appl. Phys. Lett. **53**, 791 (1988)
[4]    J. Oswald, B. Tranta, M. Pippan and G. Bauer, J. Opt. Quantum Electron. **22** 243 (1990)
[5]    J. Oswald and M. Pippan, Semicond. Sci. Technol. **8** 435 (1993)
[6]    H. Burkhard, G. Bauer and W. Zawadzki, Phys. Rev. B **19** 5149 (1979)
[7]    J. Oswald, P. Pichler, B.B. Goldberg and G. Bauer, Phys. Rev. B **49** 17029 (1994)
[8]    J. Oswald, M. Pippan, G. Heigl, G. Span, T. Stellberger, Proc. 7th Int. Conf. on NGS, Santa Fe, 1995, Inst.Phys.Conf.Ser. No 144, 155 (1995) (cond-mat/9707157)
[9]    J. Oswald, B.B. Goldberg, G. Bauer and P.J. Stiles , Phys. Rev. B **40** 3032 (1989)
[10]   A.K. Geim, P.C. Main, P.H. Beton, P. Streda and L. Eaves, Phys. Rev. Lett. **67**, 3014 (1991)